\documentclass[bibyear]{aa} 

\usepackage{graphicx}
\usepackage{txfonts}
\begin{document}

   \title{Disk stars in the Milky Way detected beyond 25 kpc from its center}
   \subtitle{}

   \author{M. L\'opez-Corredoira\inst{1,2}, C. Allende Prieto\inst{1,2},
F. Garz\'on\inst{1,2}, H. Wang\inst{3,4}, C. Liu\inst{3,4}, L. Deng\inst{3,4}}

   \institute{$^1$ Instituto de Astrof\'\i sica de Canarias, 
E-38205 La Laguna, Tenerife, Spain\\
$^2$ Departamento de Astrof\'\i sica, Universidad de La Laguna,
E-38206 La Laguna, Tenerife, Spain \\
$^3$ Key Laboratory of Optical Astronomy, National Astronomical Observatories,
Chinese Academy of Sciences, Beijing 100012, China\\
$^4$ University of Chinese Academy of Sciences, Beijing, 100012, China}

   \date{Received xxxx; accepted xxxx}

 
  \abstract
  {The maximum size of the Galactic stellar disk is not yet known.
Some studies have suggested an abrupt drop-off of the stellar density of the disk at 
Galactocentric distances $R\gtrsim 15$ kpc, which means that in practice no disk stars or only very few of them should be found beyond this limit. However, stars in the Milky Way plane are detected at larger distances. In addition to the halo component, star counts have placed the
end of the disk beyond 20 kpc, although this has not been spectroscopically confirmed so far.}
   {Here, we aim to spectroscopically confirm the presence of the disk stars up to 
much larger distances.}
   {With data from the LAMOST and SDSS-APOGEE 
spectroscopic surveys, 
we statistically derived the maximum distance at which the
metallicity distribution of stars in the Galactic plane is distinct from that of the halo populations.}
   {Our analysis reveals the presence of disk stars at $R>26$ kpc 
(99.7\% C.L.) and even at $R>31$ kpc (95.4\% C.L.).}
   {}

   \keywords{Galaxy: structure -- Galaxy: disk -- Galaxy: abundances}

\titlerunning{Disk stars at $R>25$ kpc}
\authorrunning{L\'opez-Corredoira et al.}

   \maketitle
%

\section{Introduction}

The disk of our Galaxy has an exponential radial profile 
(de Vaucouleurs \& Pence 1978; Bahcall \& Soneira 1980), which means that the density of stars quickly decreases away from the
center, although in principle a
few stars should be present at very large distances from the center and some of
them could be detected.
With a typical scale length of 2 kpc (L\'opez-Corredoira \& Molg\'o 2014; hereafter LM14)
and a solar neighborhood surface density of visible stars (main-sequence and giants) of 27 M$_\odot $pc$^{-2}$ (McKee et al. 2015), the surface density at a Galactocentric distance of $R=25$ kpc would be $\sim 5\times 10^3$ M$_\odot $kpc$^{-2}$. Only 2\% of the mass is due to giant stars (McKee et al. 2015), which are bright enough to be detected spectroscopically at these distances, so that the mass density associated with the giants would be $\sim 100$ M$_\odot $kpc$^{-2}$, that is, only very few giant stars, but a significant number are expected to be detected.
 
Some authors (Freudenreich et al. 1994; Ruphy et al. 1996; Porcel et al. 1997; Sale et al. 2010; Minniti et al. 2011; Am\^ores et al. 2017) have argued that the density of disk stars at $R>13-16$ kpc is dramatically reduced with respect to an extrapolation of the exponential disk with the scale length of the inner disk.
However, it is suspected that they expected a significant drop-off of stars because the flare of the Galactic disk becomes strong at these Galactocentric distances (LM14), and the stars are therefore
distributed over a much wider range of heights, which produces this apparent depletion of in-plane stars. 
The surface density may not fall off abruptly, but the stars would simply be redistributed at greater heights from the plane.
The flare has also been confirmed kinematically with the measured thickening of the vertical velocity distribution (Wang et al. 2017).

Momany et al. (2006) and Reyl\'e et al. (2009) investigated the outer disk, but limited to $R<20$ kpc and with large uncertainties beyond 15 kpc.
Carraro et al. (2010) found some young stars between 15 and 20 kpc from the
Galactic center. Feast et al. (2014) speculated about the interpretation
of five Cepheids in the outer disk 1-2 kpc from the plane, but their results are puzzling  since the very young population ($\sim 100$ Myr) of Cepheids
typical of a spiral arm should not be as farther away from $z=0$ plane.
Liu et al. (2017) reported that
the disk seems to extend at least up to $R=19$ kpc, and that
beyond this radius the
disk smoothly transitions to the halo without any truncation, break, or upward bending. These are further indications that the disk may not end at least out to $R=20$ kpc, but what happens beyond this distance? Can we
provide proof of the existence of disk stars farther away? The purpose of this paper is precisely answering this question.
Certainly, there are stars beyond $R=20$ kpc, but many of them belong to
the old population of the halo (Xu et al. 2017). LM14 previously
showed the existence of stars out to $R=30$ kpc, although only in regions far from the plane, and without a spectroscopic classification of their age or metallicity.  

\section{Method}

Our method in this paper is to search for a population typical of the thin disk, with a distribution of metallicities distinct from that of the halo,
shifted toward higher metallicities.
The halo metallicity distribution function (MDF) peaks at around [Fe/H]=-1.6 (Beers \& Christlieb 2005; Allende Prieto et al. 2014), while the MDF of the disk reaches its maximum between roughly -0.7 and +0.25, depending upon the height over the midplane and the radial distance. (Hayden et al. 2015). Hence, metallicity by itself is useful to separate between halo and disk populations.

Specifically, our method consists of comparing the distribution of metallicities in two samples that satisfy i) a Galactocentric distance between $R_1$ and $R_1+\Delta R$, $|z|<5$ kpc and ii) a Galactocentric distance between $R_2$ and $R_2+\Delta R$, $|z|\ge 5$ kpc. The reason we chose a height of $z=5$ kpc for the separation of the two subsamples is that 
the scale height of the thick disk is approximately 1 kpc and the flare of the outer disk can reach a thickness of several kpc for the disk (LM14). We set a fixed value of $R_1$ and $\Delta R$ and fit the value of $R_2$ in order to obtain the
same average spherical Galactocentric distance for both distributions:
$\langle r_1\rangle \approx \langle r_2\rangle $. This avoids the possible variation of the metallicity due to a gradient in the halo ([Fe/H] slightly depends on the spherical Galactocentric distance $r$; using data fom Fern\'andez-Alvar et al. (2015, Fig. 6b), we
derive a mean $\frac{d[Fe/H]}{dr}=-0.0121\pm 0.0013$ 
kpc$^{-1}$). The non-sphericity of the halo is negligible at large
radii (Xu et al. 2017).

Comparing distributions of heliocentric radial velocities might be another way of distinguishing halo and disk populations, but this is not so straightforward and would need a priori kinematic models to separate the contribution of
different Galactocentric velocity components. We therefore do not use it here.

\section{Data}
\label{.data}

We carried out our analysis with data on K-giants from LAMOST-DR3 (Liu et al. 2017) in the
optical and SDSS-APOGEE-DR14 (Majewski et al. 2017) in the near-infrared. 

The LAMOST DR3 catalog contains 5\,756\,075 spectra, for which the LAMOST pipeline has provided the metallicity [Fe/H], and the
distances were estimated from a Bayesian approach (Carlin et al. 2015) with uncertainties of about 20\%. About 70\,000 K-giants were selected from LAMOST DR3 according to the criterion of Liu et al. (2014).

The Apache Point Galactic Evolution Experiment (APOGEE) DR14 (Abolfathi et al. 2017) includes millions of spectra for
approximately 263,000 stars. Distances for the stars have been estimated with four different
methods (Schultheis et al. 2014; Santiago et al. 2016; Wang et al. 2016; Holtzman et al. 2018) and were included in a value-added catalog released in conjunction with DR14. 
The agreement among the four codes is fair, typically
within 20\%. We have adopted for our analysis the average values of the available
estimates, as well as the overall metallicity [M/H] values derived by the APOGEE
ASPCAP pipeline (Garc\'{\i}a P\'erez et al. 2016), which for DR14 have been calibrated
to match optical iron abundances ([Fe/H]) for clusters in the literature.

The spatial distribution of the stars considered here spans a range of Galactic longitudes that is accessible from
observatories in the northern hemisphere, with 
those at larger $R$ toward the anticenter. Although the overdensity of stars at $R\approx 20$ kpc 
was attributed by some authors to tidal debris of a dwarf galaxy (Monoceros Ring), 
LM14 have shown that this
hypothesis is unnecessary and that the overdensity can be explained by a flared disk.
Here we follow the argument of LM14.

\section{Results}

\begin{figure}
\vspace{0cm}
\centering
\includegraphics[width=9cm]{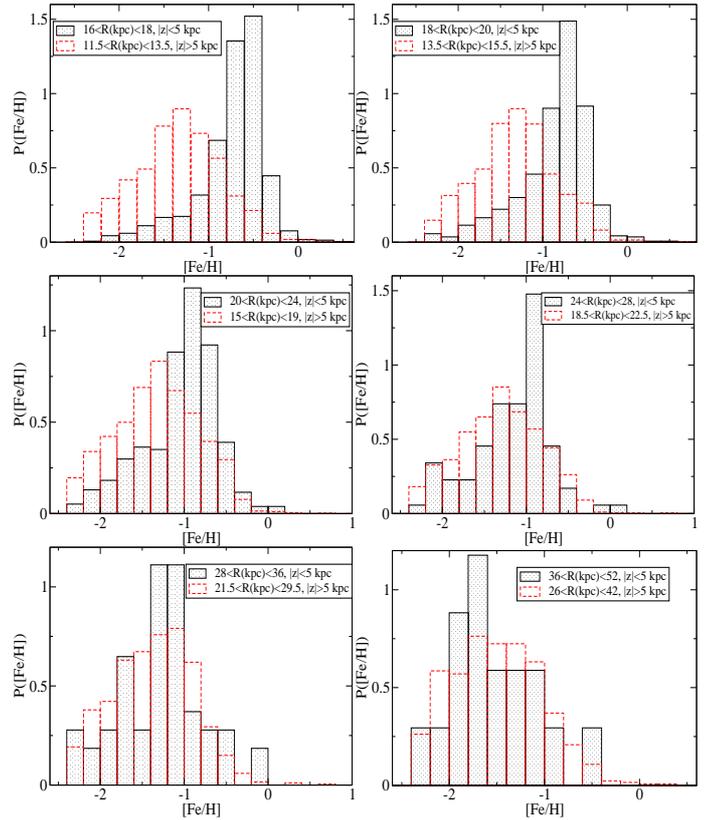}
\caption{Metallicity distributions [Fe/H] for different subsamples
of LAMOST-DR3 K giants with $R_1<R<R_1+\Delta R$, $|z|<5$ kpc, and $R_2<R<R_2+\Delta R$, $|z|\ge 5$ kpc, respectively, such that $\langle r_1\rangle \approx \langle r_2\rangle $.
Normalized such that $\int d[Fe/H]P([Fe/H])=1$.}
\label{Fig:distmetLAMOST}
\end{figure}

\begin{figure}
\vspace{0cm}
\centering
\includegraphics[width=9cm]{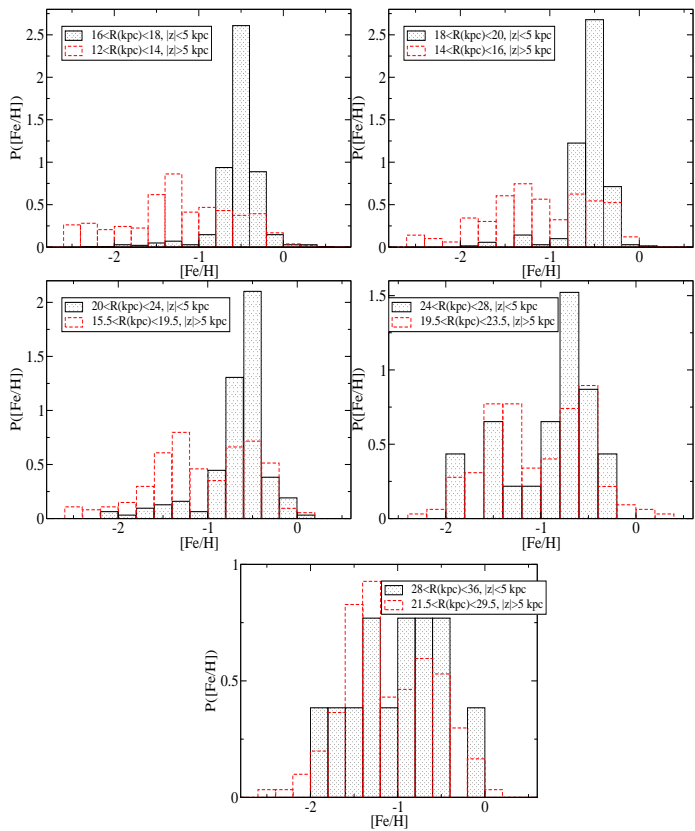}
\caption{Same as Fig. \ref{Fig:distmetLAMOST} for SDSS-APOGEE-DR14 stars.}
\label{Fig:distmetAPOGEE}
\end{figure}

The metallicity distributions for different $R_1$ and for both surveys are given in Figs. \ref{Fig:distmetLAMOST} and \ref{Fig:distmetAPOGEE}.
Possible selection effects on completeness do not affect the metallicity distribution
(Nandakumar et al. 2017).
A similar histogram was produced in Fig. 11 of Carlin et al. (2015) with LAMOST, but only with $R<20$ kpc in the plane, whereas here we analyze the distributions beyond that limit. The metallicity of LAMOST halo stars was also analyzed (Xu et al. 2017), but without
the stars in the plane with [Fe/H]$>-1$ that we include here.
Tables \ref{Tab:distmetLAMOST} and \ref{Tab:distmetAPOGEE} give the parameters of these distributions.
The disk metallicity distribution peaks between -1.0 and -0.5, whereas
the halo mean metallicity is a wider distribution with a maximum at about -1.5.
The first range is expected from an extrapolation of the
metallicity gradient from the inner disk, including both thin
and thick disks 
(Besan\c con model simulation in L\'opez-Corredoira et al. 2007, Fig. 3).

\begin{table}
\caption{Parameters in the comparison of
metallicity distributions [Fe/H] for different subsamples
of LAMOST-DR3 giants with $R_1<R<R_1+\Delta R$, $|z|<5$ kpc, and $R_2<R<R_2+\Delta R$, $|z|\ge 5$ kpc, respectively, such that $\langle r_1\rangle \approx \langle r_2\rangle $. The first three columms indicate the Galactocentric radial range, columns 4 and 5 give the number of stars, column 6 gives the average signal-to-noise ratio of the first subsample, column 7 gives the average error of the [Fe/H] measurement in the first
subsample, and column 8 gives the probability derived from a
Kolmogorov-Smirnov test
that the two metallicity distributions are identical within the random fluctuations.  The distances are listed in kpc.}
\begin{center}
\begin{tabular}{cccccccc}
$R_1$ & $\Delta R$  & $R_2$ & $N_1$ & $N_2$ &
$\langle S/N\rangle _1$  &  $\langle \Delta {\rm met}\rangle _1$ &
$P_{K-S}$ \\ \hline
16 & 2 & 11.5 & 2160 & 1850 & 18.8 & 0.23 & 1.9E-298   \\
18 & 2 & 13.5 & 699 & 1352 & 17.3 & 0.25 & 1.0E-99 \\
20 & 4 & 15.0 & 359 & 1813 & 15.5 & 0.25 & 1.6E-27 \\
24 & 4 & 18.5 & 88 & 992 & 14.2 & 0.26 & 0.014  \\
28 & 8 & 21.5 & 54 & 936 & 12.6 & 0.24 & $>0.32$ \\
36 & 16 & 26.0 & 17 & 649 & 11.0 & 0.28 & $>0.32$ \\ \hline
\label{Tab:distmetLAMOST}
\end{tabular}
\end{center}
\end{table}

\begin{table}
\caption{Same as Table \ref{Tab:distmetLAMOST} for SDSS-APOGEE-DR14 stars.}
\begin{center}
\begin{tabular}{cccccccc}
$R_1$ & $\Delta R$ & $R_2$  & $N_1$ & $N_2$ &
$\langle S/N\rangle _1$  &  $\langle \Delta {\rm met}\rangle _1$ &
$P_{K-S}$ \\ \hline
16 & 2 & 12.0 & 715 & 267 & 180.2 & 0.03 & 4.0E-82 \\
18 & 2 & 14.0 & 351 & 248 & 165.9 & 0.03 & 6.9E-38 \\
20 & 4 & 15.5 & 157 & 370 & 148.7 & 0.03 & 1.1E-17 \\
24 & 4 & 19.5 & 23 & 162 & 169.8 & 0.04 & 2.0E-3 \\
28 & 8 & 21.5 & 13 & 151 & 123.0 & 0.04 & $>0.32$ \\ 
36 & 16 & -- & 0 & -- & -- & -- & -- \\ \hline
\label{Tab:distmetAPOGEE}
\end{tabular}
\end{center}
\end{table}

The results are quite clear: significant differences are found
for $R<24$ kpc between the distributions in-plane and off-plane. The
in-plane subsamples have disk and halo stars, whereas the off-plane subsample is 
composed of halo stars alone. 
No differences are found for $R>28$ kpc in the plots.

\begin{figure}
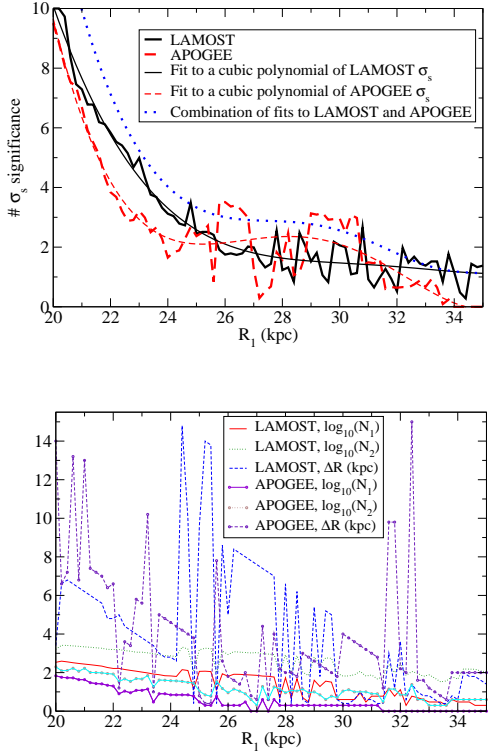

\vspace{0.4cm}
\centering
\includegraphics[width=6.3cm]{fort91.eps}\\
\vspace{.8cm}
\hspace{4mm}\includegraphics[width=6.0cm]{fort91b.eps}
\caption{Top: Maximum number (choosing the value of $\Delta R$ that gives
the maximum value) of sigmas of significance detection of
different metallicity distributions in the subsamples
with $R>R_1$, $|z|<5$ kpc, and $R>R_2$, $|z|\ge 5$ kpc, respectively, such that $\langle r_1\rangle \approx \langle r_2\rangle $.
Bottom: Parameters $N_1$, $N_2$, and $\Delta R$ corresponding to that detection.}
\label{Fig:sigmas}
\end{figure}

The significance of the distributions was evaluated with a Kolmogorov-Smirnov (K-S) test, 
which is non-parametric and distribution independent.
The K-S test has some limitations 
(Feigelson \& Babu 2012), for instance, if the model that is compared with a data set was derived from the same data set, or when two distributions
derived from data are not totally independent, but this is not the case here.
Tables \ref{Tab:distmetLAMOST} and \ref{Tab:distmetAPOGEE} give
the probability assigned by this test to explain the different distributions as due to random fluctuations. 
Errors in [Fe/H] will decrease the K-S maximum distance $D_{\rm max}$ between the two distributions, thus increasing the probability $P_{K-S}$, so they cannot be responsible for a significant detection.
When we vary $R_1$ and $\Delta R$ (in a range between 0.2 and 15.0 kpc) as free parameters, the maximum significance expressed in the equivalent number of sigmas for a given probability
(assuming a normal distribution; i.e., 1$\sigma $ is $P_{K-S}=0.317$,
2$\sigma $ is $P_{K-S}=0.0455$, 3$\sigma $ is $P_{K-S}=2.70\times 10^{-3}$, etc.) is given in Fig. \ref{Fig:sigmas}. We can account for
the effect of having a higher significance due to exploration of 
the values of $R_1$ if we fit a smooth function to the inferred significances, as done in Fig. \ref{Fig:sigmas}. A cubic polynomial fitting 
(lower order polynomials do not yield a good fit) of the significance, results in a difference with halo stars at 99.73\% C.L. (3$\sigma $) at $R_1=24.4$ kpc (LAMOST), $R_1=22.9$ kpc (APOGEE). A 5$\sigma $ detection is found at
$R_1=22.5$ kpc (LAMOST), $R_1=21.6$ kpc (APOGEE). A tentative detection at 2 $\sigma $ is for $R_1=26.3$ kpc (LAMOST), $R_1=30.3$ kpc (APOGEE).

Both surveys independently yield the same results. 
The APOGEE data show a  metallicity distribution for the
disk that is narrower than in the LAMOST data, possibly due to the lower errors in
the distance determination. The APOGEE spectra have a higher
resolution and signal-to-noise ratio than the LAMOST observations,
but a smaller spectral range. The off-plane stars in APOGEE present
a double peak, whereas in LAMOST it has only one peak, possibly due
to some miscalculation of disk stars for a great distance.
The minimum detected metallicity in LAMOST 
is also limited at [Fe/H]=-2.5 (Carlin et al. 2015), but this difference in the histograms is not important because very few stars have metallicities lower than this limit.
The remaining features in the distributions are equivalent. 

The two surveys are independent because they have only very few sources in common: of 3393 sources in the LAMOST sample with $R>16$ kpc, $|z|<5$ kpc, only 96 were observed by APOGEE, which
is a coincidence lower than 3\%. These 96 stars have similar distance estimates and metallicities, which corroborates the reliability of their determinations. In addition, the surveys operate at different wavelengths, with different instruments and analysis pipelines. Therefore, we can combine the statistics of
the two surveys as if they were independent: summing quadratically the number
of sigmas of both surveys. This is also shown in Fig. \ref{Fig:sigmas}.
This global analysis of the two surveys shows that the significant detection of a difference in metallicity distribution in in-plane stars with respect to pure halo stars, interpreted as the presence of disk stars added to the halo sources, is given for
$R>23.2$ kpc at 5$\sigma $, $R>26.0$ kpc at 3$\sigma $, and $R>31.5$ kpc at 2$\sigma $.

For LAMOST, stars with $\langle R\rangle=26.0$ kpc ($3\sigma $ global detection) or $\langle R\rangle=31.5$ kpc ($2\sigma $ global detection) have a mean heliocentric distance of 18.3 kpc and 23.9 kpc, respectively; according to the analysis of Carlin et al. (2015, Fig. 2/right), in comparison with the Besan\c{c}on model, the corresponding average systematic overestimation of distance is +2.2 and +3.2 kpc, respectively.
Wang et al. (2017, appendix) claimed that the errors of the distances given by Carlin et al. (2015) are overestimated by a factor of two.
Nonetheless, this excess of 1-3 kpc mainly affects stars with very low metallicity
stars because the authors used isochrones with solar [$\alpha $/Fe] 
(Carlin et al. 2015) and this should not affect the disk stars we
analyzed at $|z|<5$ kpc to find a distinction with halo population.
The limit of the detection of disk stars should therefore not be significantly 
affected.
For APOGEE, the distances were determined as the average of four independent methods that
were compatible with each other within the errors (see \S \ref{.data}), and the systematic
error of heliocentric distances in comparison with cluster distances is underestimated by 4\% (Wang et al. 2016). This is an average underestimation of -0.7 and -1.0 kpc for
$\langle R\rangle=26.0$ kpc ($3\sigma $ global
 detection) or $\langle R\rangle=31.5$ kpc, respectively, which
places the stars even slightly farther away.

In order to further determine possible systematic errors, we excluded from our sample
of APOGEE the stars with ASPCAP pipeline (Garc\'\i a P\'erez et al. 2016) 
flags (ASPCAPFLAG), which is a warning of some possible difficulties for an analysis of the star. This reduces the sample by 16\% of sources (22 in-plane stars at $R>24$ kpc instead of 37 without
the cut), and the radius at which
there is a $3\sigma $ detection is $R=22.5$ kpc (instead of $R=22.9$ kpc).
If we furthermore add another constraint and also remove stars with a warning flag in the
parameter STARFLAG, which is related to issues with the spectrum,
the number
of sources is reduced by 54\% with respect to the total sample
(10 in-plane stars at $R>24$ kpc instead of 37 without
the cut), and the   radius for a $3\sigma $ detection is $R=21.8$ kpc (instead of $R=22.9$ kpc).
When this last subsample is combined with only 46\% of the sources in APOGEE with LAMOST, we
find the presence of disk stars distinct from the halo sources, at $R>22.8$ kpc at 5$\sigma $, $R>24.7$ kpc at 3$\sigma, $ and $R>27.1$ kpc at 2$\sigma $.
This slight reduction of the maximum radius of the disk is due to 
the reduction of the number of sources, which makes the detection less significant 
at a given radius. We may then conclude that our results are not importantly affected
by possible misclassified sources, which should introduce noise rather than signal.

This analysis corroborates through statistical spectroscopy the lack of a radial truncation in the stellar disk observed through the fit of star counts out to 30 kpc 
(LM14).
An exponential distribution is also observed for the gas density of the Milky Way without any truncation up to a distance of 40 kpc from the center (Kalberla \& Dedes 2008).
This does not mean that radial truncations are not possible in spiral galaxies: there are other galaxies in which they are observed (van der Kruit \& Searle 1981; Pohlen et al. 2000), but the Milky Way is not one of them.

\begin{acknowledgements}
MLC and FGL were supported by the grant AYA2015-66506-P of the Spanish Ministry of Economy and Competitiveness (MINECO). Thanks are given to the anonymous referee for helpful comments and Astrid Peter (language editor of A\&A) for the revision of this paper.
Guoshoujing Telescope (the Large Sky Area Multi-Object Fiber Spectroscopic Telescope LAMOST) is a National Major Scientific Project built by the Chinese Academy of Sciences. Funding for the project has been provided by the National Development and Reform Commission. LAMOST is operated and managed by the National Astronomical Observatories, Chinese Academy of Sciences.
Funding for the Sloan Digital Sky Survey IV has been provided by the Alfred P. Sloan Foundation, the U.S. Department of Energy Office of Science, and the Participating Institutions. SDSS-IV acknowledges
support and resources from the Center for High-Performance Computing at
the University of Utah. The SDSS web site is www.sdss.org.
SDSS-IV is managed by the Astrophysical Research Consortium for the 
Participating Institutions of the SDSS Collaboration including the 
Brazilian Participation Group, the Carnegie Institution for Science, 
Carnegie Mellon University, the Chilean Participation Group, the French Participation Group, Harvard-Smithsonian Center for Astrophysics, 
Instituto de Astrof\'isica de Canarias, The Johns Hopkins University, 
Kavli Institute for the Physics and Mathematics of the Universe (IPMU) / 
University of Tokyo, Lawrence Berkeley National Laboratory, 
Leibniz Institut f\"ur Astrophysik Potsdam (AIP),  
Max-Planck-Institut f\"ur Astronomie (MPIA Heidelberg), 
Max-Planck-Institut f\"ur Astrophysik (MPA Garching), 
Max-Planck-Institut f\"ur Extraterrestrische Physik (MPE), 
National Astronomical Observatories of China, New Mexico State University, 
New York University, University of Notre Dame, 
Observat\'ario Nacional / MCTI, The Ohio State University, 
Pennsylvania State University, Shanghai Astronomical Observatory, 
United Kingdom Participation Group,
Universidad Nacional Aut\'onoma de M\'exico, University of Arizona, 
University of Colorado Boulder, University of Oxford, University of Portsmouth, 
University of Utah, University of Virginia, University of Washington, University of Wisconsin, 
Vanderbilt University, and Yale University.

\end{acknowledgements}

\end{document}